\documentstyle[preprint,aps,floats,epsfig]{revtex}
\tightenlines
\begin{document}
\title{
Directed flow of neutral strange particles at AGS
}
\bigskip
\author{
Bin Zhang\footnote{Email: bzhang@kopc1.tamu.edu}$^{\rm a}$,
C.M. Ko\footnote{Email: ko@comp.tamu.edu}$^{\rm a}$,
Bao-An Li\footnote{Email: bali@navajo.astate.edu}$^{\rm b}$,
and Andrew T. Sustich\footnote{Email: sustich@navajo.astate.edu}$^{\rm b}$}
\address{$^{\rm a}$Cyclotron Institute and Physics Department,\\
Texas A\&M University, College Station, TX 77843-3366, USA}
\address{$^{\rm b}$Department of Chemistry and Physics\\
Arkansas State University, P.O. Box 419\\
State University, AR 72467-0419, USA}
\maketitle

\begin{abstract}
Directed flow of neutral strange particles in heavy ion collisions at AGS 
is studied in the ART transport model. Using a lambda mean-field 
potential which is 2/3 of that for a nucleon as predicted by the 
constituent quark model, lambdas are found to flow with protons but 
with a smaller flow parameter as observed in experiments. For kaons, 
their repulsive potential, which is calculated from the impulse 
approximation using the measured kaon-nucleon scattering length, 
leads to a smaller anti-flow than that shown in
the preliminary E895 data.  Implications of this discrepancy are
discussed.
\bigskip

\noindent{PACS number(s): 25.75.Ld, 13.75.Jz, 21.65.+f}
\end{abstract}
\bigskip

Studies of kaon collective flow in heavy ion collisions have
been very useful in extracting the kaon properties in hot dense matter
\cite{rev}. Such knowledge is important in 
understanding the possibility of kaon condensation in the core 
of neutron stars \cite{gbrown1,vthorsson1} and the existence of 
mini black holes \cite{gbrown2,gbrown3}. Available experimental data 
from both the FOPI \cite{jritman1,dbest1} and the KaoS collaboration 
\cite{shin} at SIS/GSI show a negligible kaon flow in collisions of Ni+Ni at 
$E_{\rm beam}/{\rm nucleon}=1.93$ GeV and Au+Au at 1 GeV, respectively. 
Based on transport models, it has been shown that a vanishing kaon flow is 
consistent with a repulsive kaon potential in medium 
\cite{gqli4,zswang1,ebratkovskaya1,cfuchs1,gqli1}. 
For heavy ion collisions at higher energies available from the AGS, 
the directed flow of neutral strange particles ($K^0$ and $\Lambda$)
is being studied experimentally by the E895 collaboration \cite{pchung1}. 
Preliminary data have shown that lambdas flow in the same 
direction as protons but with a smaller flow parameter, and 
neutral kaons have a strong negative flow relative to that of nucleons.
To understand these experimental observations, we have studied 
using A Relativistic Transport (ART) model \cite{bali1} the
directed flow of these neutral strange particles in Au on Au collisions
at a beam momentum of 6 GeV/c and an impact parameter of 4 fm.
We note that particle directed flow is characterized by the average 
in-plane transverse momentum as a function of rapidity.
In the central rapidity region, one can also introduce the flow 
parameter which measures the change of average in-plane 
transverse momentum with respect to the rapidity. 

\bigskip
\begin{figure}[ht]
\centerline{\epsfig{file=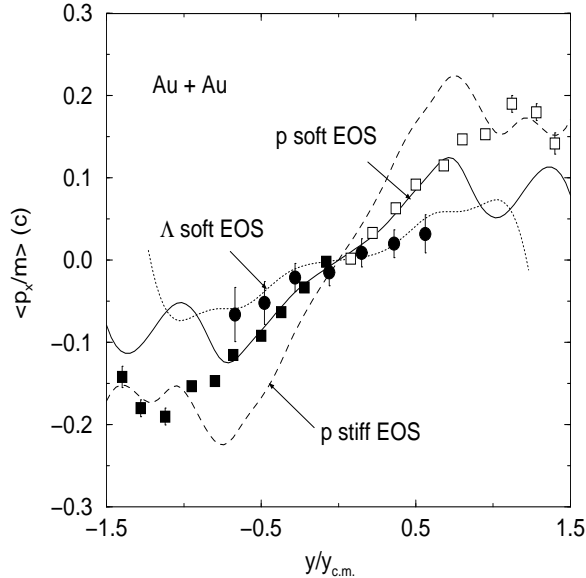,height=3.0in,width=3in,angle=-90}}
\vspace{0.3in}
\caption{Proton and lambda directed flow from Au+Au collisions at a beam 
momentum of 6 GeV/c per nucleon and an impact parameter of 4 fm. 
Results from the ART model are shown by lines. The preliminary E895 data
are open squares for the proton flow and solid circles for
the lambda flow. Filled squares are the reflected proton data.}
\label{fig1}
\end{figure}

For the lambda potential, we take it to be 2/3 of that of protons. 
This follows from the light quark contents of proton and lambda,
and is also consistent
with that determined from the structure of hypernuclei \cite{dmillener1}.
The nucleon potential used in the ART model is the standard Skyrme-type
derived from either a soft nuclear equation of state with 
a compressibility of 200 MeV or a stiff nuclear equation of state
with a compressibility of 380 MeV.
In Fig. 1, we show the results for the proton and lambda directed flow
from the ART model together with the preliminary E895 data. The latter
are based on the analysis of 8500 semi-central
Au(6 AGeV)+Au events. To improve the statistics, the ART result for 
a given rapidity has been obtained by averaging the values  
at both positive and negative rapidities. Typical statistical errors
in the ART results for the proton flow are around 
$0.1\%$ and for lambda flow are around $0.3\%$.  In our calculations, 
we have neglected the effect of rescatterings among lambdas. This
should be a good approximation as 
the lambda to nucleon ratio is only about $1\%$.
We find that at central rapidities, the soft nuclear equation of state 
gives a better description of the measured proton directed flow, 
while at high rapidities ($y/y_{c.m.}>0.7$), 
the stiff nuclear equation of state describes the data better.
For the lambda flow, the ART model gives a smaller value
at central rapidities than that for protons. This result is similar to 
that obtained in heavy ion collisions at lower energies of about 2 GeV/nucleon
\cite{liko96}. As shown in Fig. 1, using the soft equation of state
can indeed account for the experimentally
observed weaker directed flow of lambdas. 

In the ART model, the kaon mean-field potential is obtained 
from the impulse approximation. In this approximation, the
kaon dispersion relation is related to the kaon-nucleon scattering length 
via
\begin{equation}
\omega(\vec{k}, \rho)=\left[m_K^2+\vec{k}^2-4\pi\left(1+\frac{m_K}{m_N}
\right)a_{KN}\rho\right]^{1/2},
\label{eq:impulseapp}
\end{equation}
where $a_{KN}\approx -0.255$ fm is the isospin-averaged kaon-nucleon
scattering length, $m_K$ and $m_N$ are the kaon and nucleon masses,
and $\rho$ is the local baryon density.
The kaon mean-field potential is then defined as
\begin{equation}
U(\vec{k},\rho)=\omega(\vec{k}, \rho)-(m_K^2+\vec{k}^2)^{1/2}.
\end{equation}
Because of the repulsive nature of the kaon potential, the predicted 
directed flow for $K^0$'s is opposite from that of protons and has
almost zero value at the central rapidity.
However, the charge-exchange reactions between
$K^0$ and $K^+$ have not been included in the ART model. 
To obtain an upper limit on the effect of charge-exchange reactions, 
we have included the Coulomb interaction also for $K^0$'s.
Results from such a calculation are compared with the preliminary data
from E895, which are for $K^0_S$'s reconstructed from
their charged pion decays in semi-central collisions. Since
the number of $\bar{K^0}$'s is much smaller than that of $K^0$'s at AGS 
energies, the experimentally observed $K^0_S$'s thus come mostly from $K^0$'s. 

\begin{figure}[ht]
\centerline{\epsfig{file=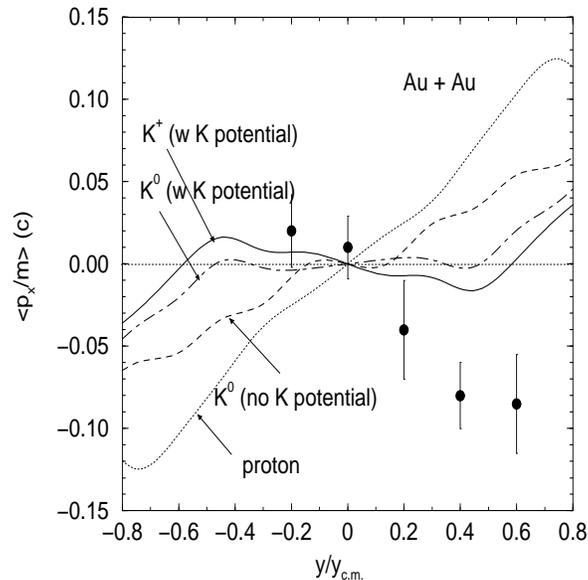,height=3in,width=3in,angle=-90}}
\vspace{0.5in}
\caption{Comparisons of the kaon flow from the ART model (curves) with
preliminary E895 data (solid circles).}
\label{fig2}
\end{figure}

In Fig. 2, we compare the kaon flow results from the ART model 
with the E895 data. Typical statistical errors
are around $1\%$ for the kaon flow from the ART model. 
It is clearly seen that without kaon mean-field potential, 
$K^0$'s flow with nucleons. The repulsive kaon mean-field
potential makes kaons flow away from nucleons. 
In addition, charge-exchange reactions enhance the anti-flow of kaons 
with respect to protons due to additional Coulomb repulsion. 
We see that even if all $K^0$'s are $K^+$'s in the reaction, $K^0$'s can 
get at most an average $p_x/m$ of about 0.01c which is almost five times 
smaller than that observed by the E895 collaboration. Our results
on $K^+$ flow are, however, consistent with the
preliminary data from the E917 collaboration at AGS \cite{ogilvie}.
They are also similar to the theoretical results for lower energies
\cite{jritman1,dbest1,zswang1,ebratkovskaya1,cfuchs1,gqli1,bali2,wreisdorf1},
i.e., kaons have essentially vanishing flow.
On the other hand, results from the RQMD model show a positive
$K^0_S$ flow due to neglect of the kaon mean-field potential
\cite{chung}.

Since kaon directed flow is affected by the kaon potential in nuclear
medium, using different potentials can give different values for the flow
parameter. Various models \cite{dkaplan1,hpolitzer1,mlutz1} 
have suggested that the kaon energy in the medium
can be written as
\begin{equation}
\omega(\vec{k},\rho)=\left[m_K^2+\vec{k}^2-a_K\rho_S+(b_K\rho)^2\right]^{1/2}
+b_K\rho.
\label{eq:elag}
\end{equation}
In the above, $\rho_S$ is the baryon scalar density, which is 
smaller than the baryon density $\rho$.  The parameter 
$b_K=3/(8f_\pi^2)=0.333$ GeV/fm$^3$ determines the repulsive 
kaon-nucleon vector interaction, while
the parameter $a_K$ is used to adjust the strength of the attractive 
scalar interaction in order to effectively take into account higher-order 
corrections to the mean-field prediction. At SIS energies, $a_K$ was found 
to be $0.22$ GeV$^2$fm$^3$ in order to reproduce the $K^+$ 
spectra \cite{gqli2}.  At normal nuclear matter density, 
Eq. (\ref{eq:elag}) gives a kaon potential of about 30 MeV 
and is consistent with that from the impulse approximation using
the empirical kaon-nucleon scattering length.
As shown in Ref. \cite{gqli4}, a large negative
kaon directed flow can be obtained when higher-order contributions cancel a 
large fraction of the scalar interaction at high densities. 
Since preliminary E895 data show a very large increase in kaon anti-flow 
compared to the SIS data \cite{jritman1} at lower beam energies, this 
abrupt change may give us new information about the kaon in-medium
properties. 

Another possible reason for the large kaon anti-flow is related to the 
kinematics of three-body phase space in kaon production from 
baryon-baryon interactions. A recent publication \cite{cdavid1} 
showed that the kinematics due to an isotropic three-body phase space 
might lead to a more isotropic kaon flow compared to that from 
the traditional cluster phase-space kinematics as used in the ART model. 
At SIS energies, this effect is not important as 
kaon production is dominated by meson-baryon interactions.
Since kaon production from baryon-baryon interactions becomes
more important at AGS energies, using the
isotropic three-body phase space, which is not included in the ART model,
is expected to make initial kaons stay away from the nucleons and thus
lead to a larger kaon anti-flow. However, this may also affect
the lambda directed flow. A detailed
study of the directed flow of strange particles in conjunction with other 
variables, e.g., the differential \cite{bali} and 
elliptic flow, and their dependence on the system size, beam energy, and 
impact parameter will help to understand the relative importance of 
these different mechanisms. 

In summary, we have studied the directed flow of neutral strange 
particles in the ART model. Compared with preliminary E895 data, 
the ART results show that the smaller lambda directed flow relative
to the proton flow can be accounted for by a weaker mean-field potential
as in the constituent quark model and from the hypernuclei phenomenology.
However, the $K^0$ flow cannot be explained by the kaon mean-field 
calculated from the impulse approximation using the 
empirical kaon-nucleon scattering length.
This may need the introduction of different density dependence of 
the kaon potential at high baryon densities than the naive linear one.

We thank D. Best and R. Lacey for helpful discussions.
This work was supported in part by NSF Grant PHY-9870038, the
Robert A. Welch Foundation Grant A-1358, and the Texas
Advanced Research Project FY97-10366-068.

\end{document}